\documentclass[aps,pra,reprint,superscriptaddress,twocolumn,showpacs,amsmath]{revtex4-1}

\usepackage{amssymb}
\usepackage{graphicx}
\usepackage{bm}
\usepackage{hyperref}

\newcommand{\sgn}{\operatorname{sgn}}
\newcommand{\rect}{\operatorname{rect}}

\begin{document}

\title{Parity breaking with a nonlinear optical double-slit configuration}

\author{Vassilis Paltoglou}
\affiliation{Department of Mathematics and Applied Mathematics, University of Crete, 70013 Heraklion, Crete, Greece}

\author{Nikolaos K. Efremidis}
\email[Corresponding author: ]{nefrem@uoc.gr}
\affiliation{Department of Mathematics and Applied Mathematics, University of Crete, 70013 Heraklion, Crete, Greece}

\date{\today}



\begin{abstract}
We consider an optical nonlinear interferometric setup based on Young's double-slit configuration where a nonlinear material is placed exactly after one of the two slits. We examine the effects of Kerr nonlinearity and multi-photon absorption in the resulting interference pattern. The presence of nonlinearity breaks the transverse spatial symmetry of the system, resulting to a modified intensity pattern at the observation plane as a function of the incident intensity. Our theoretical model, based on the modification of the optical path due to the presence of nonlinearity, is surprisingly accurate in predicting the intensity profile of the main lobes for a wide range of parameters. We discuss about potential applications of our model in nonlinear interferometry. Specifically, we show that it is possible to measure both the multi-photon and the Kerr coefficients of a nonlinear material based on the spatial translation of the interference pattern as a function of the incident intensity. 
\end{abstract}

\maketitle

\section{Introduction}

The double-slit experiment is perhaps one of the most fundamental in quantum mechanics illustrating the wave-particle duality of a quantum wavepacket~\cite{feynman-vol3-1963}. 
The setup was originally introduced by Young to demonstrate the classical wave nature of light~\cite{young-ptrs1804}. 
Experiments were originally carried out with photons~\cite{taylo-pcps1909} until 1961 when the first experiment was performed using electrons~\cite{johns-zp1961}. 
Since then several experiments have studied different aspects of electron double-slit diffraction~\cite{merli-ajp1976,tonom-ajp1989,bach-njp2013,richt-prl2015} as well as the diffraction of heavier molecules~\cite{arndt-nature1999,schmi-prl2008,eiben-pccp2013} and Bose-Einstein condensates~\cite{andre-science1997}. 
Experiments have also considered moving slits in connection to the Einstein-Bohr gedanken experiment~\cite{schmi-prl2014,liu-np2014} as well as temporal interference of pulses~\cite{linde-prl2005}.

Nonlinear extensions of the double-slit configuration have been considered in optics, in the case of self-focusing~\cite{roman-oe2006} as well as self-defocusing nonlinearities~\cite{sun-ls2008}. In these works a double-slit aperture is placed in front of the medium, and the nonlinear dynamics are recorded at the observation plane. The  double-slit experiment has also been investigated in terms of nonlinear Raman micorscopy~\cite{gache-prl2010}.
In a recent work, we have shown that by introducing a Kerr nonlinear optical material having the form of a thin slab, results to a controllable shift of the intensity pattern at the observation plane~\cite{palto-ol2015}. The main difference from the previous works~\cite{roman-oe2006,sun-ls2008} is that the nonlinear material is located in front of one of the two slits thus breaking the parity symmetry of the problem in the transverse plane.

In this paper, we consider a double slit configuration similar to that discussed in~\cite{palto-ol2015}, where, in addition to the Kerr effect, we take into account losses due to multiphoton absorption and the difference in the refractive index of the material. In the proposed system the left slit covered by the nonlinear material is having the form of a relatively thin slab. 
The parity symmetry breaking in the transverse plane due to the presence of the slab material results to a modified intensity pattern at the output as a function of the incident intensity. 
We develop a simple theoretical model that takes into account these effects in terms of modifying the effective optical path, and then utilize diffraction theory to obtain an analytic expression of the field amplitude at the observation plane. Our theoretical model is in excellent comparison with our direct numerical simulations. We consider possible applications of our system in nonlinear interferometry, and in particular in measuring optical nonlinearities~\cite{sheik-ol1989}. For such applications, we rely on measuring the spatial translation of the intensity maxima at the observation plane as a function of the incident beam intensity. We show that it is possible to measure both the Kerr nonlinearity as well as the multiphoton absorption coefficients of a given material.

\section{Modeling of the nonlinear double slit configuration}
\begin{figure}[t]
\centerline{\includegraphics[width=0.8\columnwidth]{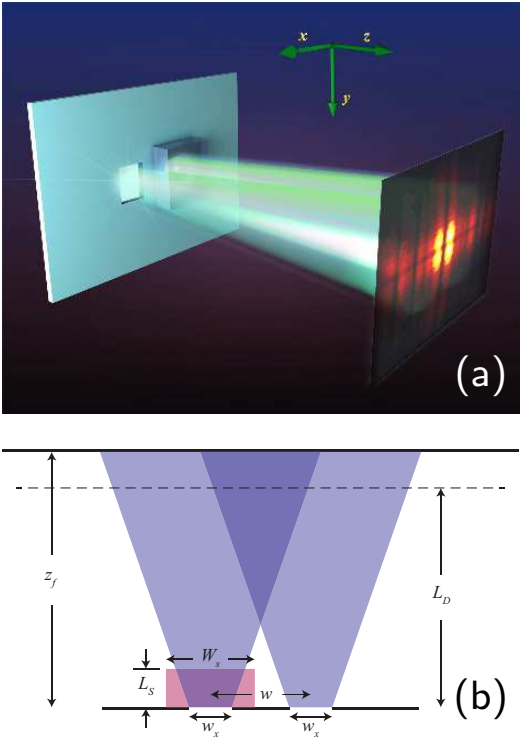}}
\caption{Three-dimensional schematic illustration (a) and its cross section (b) of the nonlinear double slit configuration. A nonlinear material having the form of a slab is placed in front of the left slit. At the observation plane the interference pattern is affected by the nonlinear properties of the slab.\label{fig:1}}
\end{figure}
Let us consider the double-slit configuration shown in Fig.~\ref{fig:1}. A coherent monochromatic laser light source with intensity $I_0$ is normally incident at the aperture plane ($z=0$). The two rectangular slits have dimensions $w_x\times w_y$ and their centers are separated by a distance $w$ along the $x$-direction. A nonlinear slab having length $L_s$ in the $z$-direction is placed in front of the left slit and completely covers it. The nonlinear effects that are taken into account are the Kerr (cubic) nonlinearity as well as losses due to multiphoton absorption. Thus the complex refractive index dependence of the material is given by 
\begin{equation}
n(I)=n_0+\gamma I+i\frac{\beta^{(K)}}{2k_0}I^{K-1}
\end{equation}
where 
$n_0$ is the linear part of the refractive index, 
$\gamma$ is the Kerr coefficient, 
$I$ is the beam intensity, 
$k_0=\omega/c$ is the free space wavenumber, 
$\omega$ is the optical frequency, 
$c$ is the speed of light, 
and $\beta^{(K)}$ is the $K$-photon absorption coefficient. 
After the beams propagate through the air and the nonlinear material, the intensity pattern of their interference is recorded at the observation plane $z=z_f$. 

The paraxial equation that describes the beam propagation in the nonlinear slab and the air is given by
\begin{equation}
i\psi_z+\frac1{2k}\nabla_{x,y}^2\psi
+\gamma k_0|\psi|^2\psi
+i\frac{\beta^{(K)}}2|\psi|^{2(K-1)}\psi
=0
\label{eq:paraxial}
\end{equation}
where $\psi$ is the field amplitude, $\nabla_{x,y}^2=\partial_x^2+\partial_y^2$, $k=n_0\omega/c=n_0k_0$ is the wavenumber inside the slab. When the beam(s) propagate in the air $k=k_0$ and the nonlinear terms are ignored in Eq.~(\ref{eq:paraxial}). At this point it is important to define the relevant length scales of the problem. 
To this end we introduce normalized coordinates: 
We scale the field to the square root of the input intensity $\psi=I_0^{1/2}\Psi$. Furthermore, we measure distances in the transverse plane according to the aperture size along the $x$-direction ($x_0=w_x$), $X=x/x_0$, $Y=y/x_0$ and scale the propagation distance according to the diffraction (Rayleigh) length in free space $Z=z/z_0$, i.e., 
\begin{equation}
z_0=L_D=k_0x_0^2.
\end{equation}
Therefore Eq.~(\ref{eq:paraxial}) in normalized dimensionless form can be expressed as
\begin{equation}
i\Psi_Z +\frac1{2n_0}\nabla_{X,Y}^2\Psi+
\Gamma\frac{L_D}{L_{NL}}|\Psi|^2\Psi
+i\frac{L_D}{2L_{MP}^{(K)}}|\Psi|^{2(K-1)}\Psi=0.
\label{eq:paraxial_normalized}
\end{equation}
where the nonlinear, and the multi-phonon (loss) lengths are defined as 
\begin{equation}
L_{NL}=\frac1{|\gamma| I_0k_0},\quad
L_{MP}^{(K)} = \frac1{\beta^{(K)}I_0^{K-1}},
\label{eq:Lnl}
\end{equation}
and $\Gamma=\sgn(\gamma)$. The length scales of the problem can be written in dimensionless form in terms of the diffraction length as $l_s=L_S/L_D$, $l_{NL}=L_{NL}/L_D$, $l_{MP}^{(K)}=L_{MP}^{(K)}/L_D$, and $Z_f=z_f/L_D$. Finally, the observation plane ($z=z_f$) can be selected to be in the near (Fresnel) field or in the far field. Note that for propagation in the air the nonlinear terms in Eq.~(\ref{eq:paraxial_normalized}) are ignored and $n_0=1$.

We need to identify the relations between the relevant length scales of the problem. First of all, our theory relies on the condition that the slab length is much smaller than the diffraction length $L_S\ll L_D$. This warranties that inside the slab diffraction is not going to play an important role. 
The above inequality becomes stronger by noting that $L_D$ is the diffraction length in free space, whereas inside the material its value increases to $n_0L_D$.
Furthermore, the distance of the observation plane $z_f$ (assuming that the slit separation is at most a few times larger than the aperture) should be larger or of the same order as the diffraction length, so that we are able to observe interference effects. More precisely, the contrast of the interference fringes (and thus the clarity of the results) increases as $z_f$ increases. We rely on nonlinear phase shift accumulation in order to observe changes in the interference pattern, as a function of the incident intensity. As the intensity increases the length scales $L_{NL}$, and $L_{MP}^{(K)}$ become smaller. In order to avoid strong nonlinear focusing as well as nonlinear instabilities it is important that the nonlinear length is, at minimum, of the same order as the slab length $L_{NL}\gtrsim L_S$. However, additional contributions from the multiphoton absorption can play an important role in suppressing such detrimental nonlinear effects provided that $L_{MP}^{(K)}$ is of the same order as $L_{NL}$. In this case, the above condition ($L_{NL}\gtrsim L_S$) can be further relaxed. 

Notice that there are two nonlinear length scales, $L_{NL}$ and $L_{MP}^{(K)}$, both of which are functions of the intensity with different exponents (except in the case of two photon absorption). In this respect it is important to identify a reference value of the nonlinear length $L_{NL,R}$ at which the ratio  $L_{NL}/L_{MP}^{(K)}$ is constant and equal to the two-photon case. This can be done by defining
\begin{equation}
\sigma^{(2)} = 
\frac{L_{NL}}{L_{MP}^{(K)}}\left(\frac{L_{NL}}{L_{NL,R}}\right)^{K-2}
\label{eq:sigma2}\end{equation}
where $\sigma^{(2)}$ is this constant ratio at the reference value. The above definition is particularly useful for comparing results with different multiphoton exponents. An additional effect that is not taken into account is that of linear losses. The reason that we choose not to include linear losses is because it adds complexity to our system (and to the resultant formulas presented below) but is only going to modify the results quantitatively and not qualitatively. However, and for the sake of completeness,  analytical expressions in the case of linear losses are presented in the Appendix.

\section{Asymptotics}

Assuming that the two apertures are equally and evenly illuminated leads to the following initial condition
\[
\psi(z=0)=
\sqrt{I_0}\left[\rect\frac{x+w/2}{w_x}+\rect\frac{x-w/2}{w_x}\right]
\rect\frac y{w_y}
\]
or, in normalized coordinates,
\begin{multline*}
\Psi(Z=0)=
\left[\rect \left(X+\frac w{2w_x}\right)+
\rect \left(X-\frac w{2w_x}\right)
\right]\times \\
\rect\frac {Yw_x}{w_y}.
\end{multline*}
Due to reflections, the intensity transmitted from the air to the slab and subsequently from the slab to the air can be computed by utilizing the Fresnel formulas.
However, for simplicity of our results, we consider that the transmission is perfect -- a condition that can be fulfilled with antireflective coatings. 
Assuming that diffraction is not significant inside the slab ($L_S\ll L_D$), we can find a simplified expression for the dynamics of the field that passes through the slab. This is done by directly solving Eq.~(\ref{eq:paraxial_normalized}) with constant amplitude as initial condition in the two-dimensional transverse plane. The reduced system of two ordinary differential equations can be directly integrated for the amplitude and the phase. Specifically, due to multi-photon absorption the intensity at the end of the slab $z=L_S$ reduces to 
\begin{equation}
I_{NL} = \frac{1}{\left[1+\frac{L_S}{L_{MP}^{(K)}}(K-1)\right]^{1/(K-1)}}.
\label{eq:Inl}
\end{equation}
The expression for the nonlinear phase accumulation at $z=L_S$ depends on the exponent $K$. Specifically for two photon absorption we have
\begin{equation}
\phi_{NL} = 
\frac{\Gamma L_{MP}^{(2)}}{L_{NL}}\log \left(1+\frac{L_S}{L_{MP}^{(K)}} \right)
\label{eq:phinl1}
\end{equation}
whereas for $K>2$
\begin{equation}
\phi_{NL} = 
\frac{\Gamma L_{MP}^{(K)} } {L_{NL} (K-2)} 
\left[\left(\frac{L_S}{L_{MP}^{(K)}} (K-1)+1\right)^{\frac{K-2}{K-1}}-1\right].
\label{eq:phinl2}
\end{equation}
Actually, as expected, by taking the limit $K\rightarrow2$ in Eq.~(\ref{eq:phinl2}) we obtain Eq.~(\ref{eq:phinl1}). 
Due to the higher refractive index inside the slab the left arm will accumulate a constant additional phase $\phi_L=(n_0-1)k_0L_S$. This phase is not important and thus in all our simulations we select the parameters involved so that $\phi_L$ is a multiple of $2\pi$. An additional outcome of the higher index inside the slab is that it reduces the amount of diffraction.
At this point, we need to utilize the assumptions that (i) we are working in the paraxial regime and (ii) the slab length is much smaller than the diffraction length. 
Our main outcome is that the effects of the presence of nonlinearity inside the slab [as described by Eqs.~(\ref{eq:Inl})-(\ref{eq:phinl2})] can be incorporated to the initial condition as 
\begin{multline}
\psi(z=0)=\sqrt{I_{NL}}e^{i(\phi_{NL}+i\phi_L)}\rect\frac{x+w/2}{w_x}+ \\
\sqrt{I_0}\rect\frac{x-w/2}{w_x}
\rect\frac y{w_y}.
\end{multline}
We then assume that both arms of the beam effectively propagate in the air for $z=z_f$. Utilizing the Fresnel diffraction integral we obtain
\begin{multline}
\psi=\sqrt{I_0I_{NL}}
e^{i(\Phi_{NL}+\phi_L)}
{\cal I}(x,-w/2,w_x)+ \\
\sqrt{I_0}{\cal I}(x,w/2,w_x)
{\cal I}(y,0,w_y)
\label{eq:fresnel}
\end{multline}
where
\[
{\cal I}(x,\xi_c,w_x) = 
\frac{1}{\sqrt{2i}}[F(W_+(x,\xi_c,w_x))-F(W_-(x,\xi_c,w_x))],
\]
the function $F(t)$ is the following sum of the Fresnel integrals
\[
F(t)=C(t)+iS(t) = \int_0^te^{i\pi s^2/2}ds,
\]
and 
\[
W_\pm(x,\xi_c,w_x) = \sqrt{k/(\pi z)}
\left(x+\xi_c\pm w_x/2\right).
\]
The limiting case of 1D diffraction is obtained from the above equations by taking the limit $w_y\rightarrow\infty$ resulting to ${\cal I}(y,0,w_y)\rightarrow1$. In the following simulations we are working exactly in this limit ($w_y\rightarrow\infty$) that reduces the effective dimensionality of the problem. The case of a two-dimensional square configuration in the case of Kerr nonlinearity was considered in~\cite{palto-ol2015}.

\begin{figure}
\centerline{\includegraphics[width=\columnwidth]{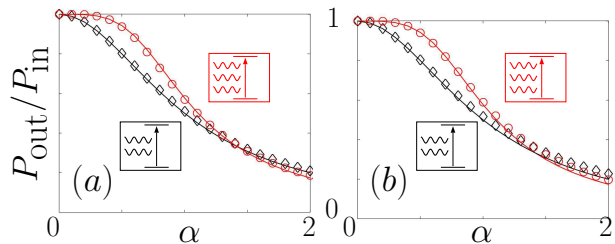}}
\caption{The fraction of the transmitted to the incident power ($P_{\mathrm{out}}/P_{\mathrm{in}}$) as a function of the relative incident amplitude $\alpha$ for a single slit configuration with a nonlinear slab placed in from of the slit. In the left and right columns the medium is self-focusing and self-defocusing, respectively. As indicated by the insets the two curves correspond to the two-photon absorption (diamonds) and the three-photon absorption (circles). In all cases $L_{NL,R}=0.025L_D$ and $\sigma^{(2)}=1$. 
 \label{fig:intensity}}
\end{figure}

\begin{figure}[t]
\centerline{
\includegraphics[width=\columnwidth]{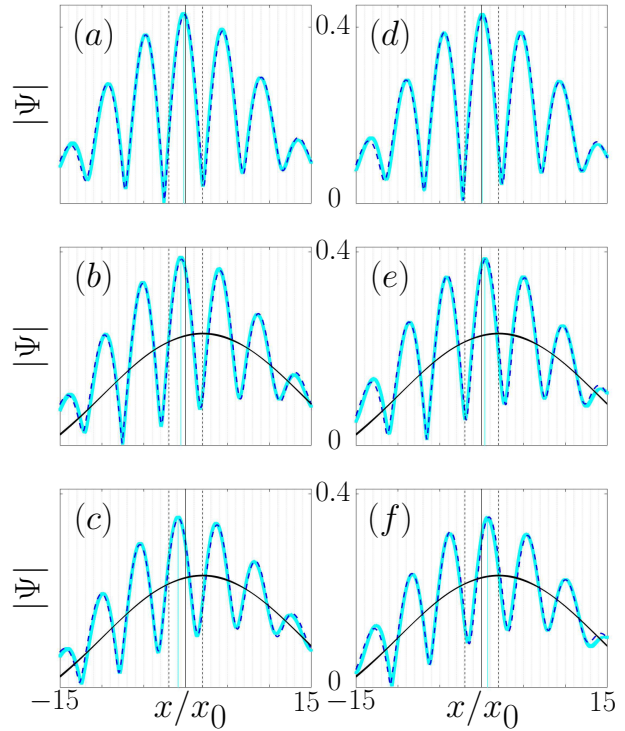}
}
\caption{
Typical amplitude interference patterns of the nonlinear double slit experiment at the observation plane $z_f=3L_D$ in the case of two-photon absorption. The slit separation is $w=4w_x$, $L_S = 0.025L_D$, $L_{NL,R}=0.025L_D$, $\sigma^{(2)}=1$. The left (right) column is obtained for self-focusing (self-defocusing) nonlinearity. The three rows (from top to bottom) correspond to relative incident amplitude $\alpha=0.5,1,1.5$, respectively. The dotted (black) curves are the numerical results while the cyan solid curves are obtained from the analytic formulas. The black solid envelope (the curve that does not exhibit oscillations) in (b)-(c) and (e)-(f) is the diffraction pattern as obtained by illuminating only the right slit.
\label{fig:IP}}
\end{figure}

\section{Numerical results}
In all our simulations we assume that the slit separation is $w=4w_x$. 
In Fig.~\ref{fig:intensity} we illuminate only the left slit and measure the ratio of the total power at the output of the slab normalized to the incident power as a function of 
\begin{equation}
\alpha=\sqrt{\frac{L_{NL,R}}{L_{NL}}}=\sqrt{\frac{l_{NL,R}}{l_{NL}}},
\end{equation}
a parameter that is proportional to the incident beam amplitude (we can refer to this parameter as the relative incident amplitude). 
Any deviations between numerical and predicted values should be attributed mainly to the effect of nonlinear diffraction inside the material which modifies the intensity pattern during propagation. As we can see in Fig.~\ref{fig:intensity} for both self-focusing and self-defocusing nonlinearity, and in both the cases of two-photon and three-photon absorption the comparison between theoretical and numerical results is excellent. 
Note that this result is somehow surprising: Although the slab length in our simulations is 1/40th of the diffraction length, diffraction effects are significant inside the slab. This is mainly due to the abrupt intensity changes of the initial condition arising from the presence of the aperture. At the output of the sample the intensity pattern does not resemble a constant function. In addition, nonlinear effects can further modify the intensity pattern at the output of the slab.
In our theoretical model the fraction of the transmitted over the incident power is equal to the fraction of the transmitted over the incident intensity ($P_\mathrm{out}/P_\mathrm{in}=I_\mathrm{out}/I_\mathrm{in}$). However, in the numerical simulations the intensity at the output is not constant in the transverse plane and thus is integrated to obtain the total power. 
It is important to note that from Eq.~(\ref{eq:Inl}) the predicted value of the intensity at the output of the slab depends only on the multiphoton absorption coefficient $l_{MP}^{(K)}$ (and thus is independent from the nonlinear length $l_{NL}$ as well as the sign of the nonlinearity). 
In the numerical results, we observe that the total power at the output is slightly higher in the self-defocusing case for large values of the incident amplitude. This happens mainly because the maximum amplitude of a modulated wave under the action of self-defocusing nonlinearity has the tendency to decrease as it propagates. Thus the amount of losses it experiences is less than those expected if the nonlinearity is self-focusing where the maximum amplitude has the tendency to increase.

\begin{figure*}
\centerline{\includegraphics[width=\textwidth]{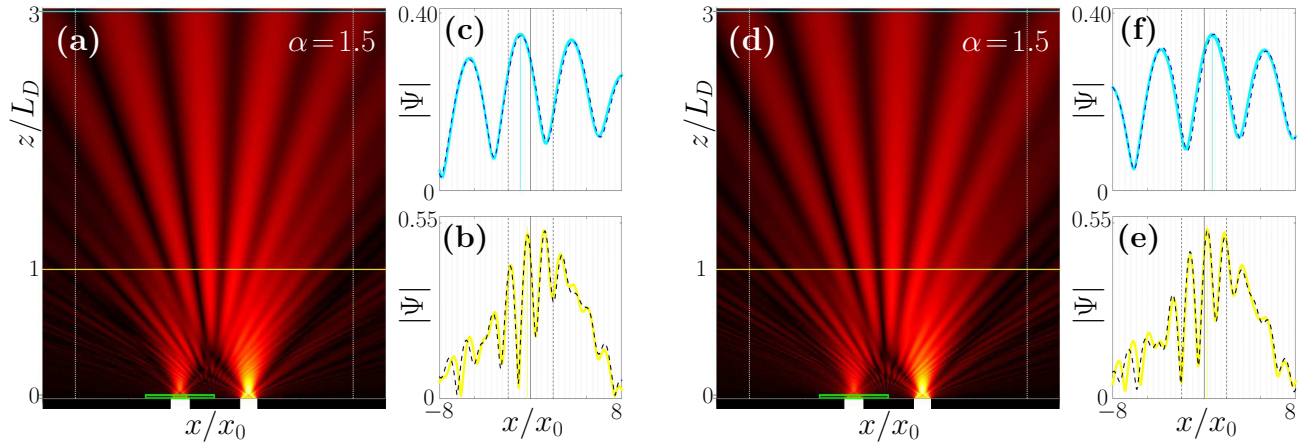}}
\caption{Dynamics of the double-slit configuration for self-focusing (a)-(c) and self-defocusing (d)-(f) nonlinearity and two-photon absorption. The geometric and material parameters are the same with those of Fig.~\ref{fig:IP} while the relative amplitude of the incident wave is $\alpha=1.5$. In (a) and (d) the amplitude dynamics are shown. In (b), (e) and (c), (f) the amplitude is depicted at the transverse plane $z=L_D$ and $z=3L_D$, respectively. In (b)-(c) and (e)-(f) the dashed curves are the numerical results whereas the solid curves are the theoretical predictions. \label{fig:dynamics}}
\end{figure*}
Typical interference patterns at $z=3L_D$ are presented in Fig.~\ref{fig:IP} for both the cases of self-focusing and self-defocusing nonlinearity. 
The comparison between the numerical results (dashed curves) with the theoretical predictions (solid curves) is excellent. 
As we can observe, by increasing the incident beam intensity the interference pattern shifts to the left (for self-focusing nonlinearity) or to the right (for self-defocusing nonlinearity). In addition, due to the presence of multi-photon absorption, the amount of intensity transmitted from the left slit is reduced. Thus for increased values of the incident beam intensity the overall envelope shifts to the right and resembles the single slit intensity pattern. For comparison, the single right slit diffraction pattern is shown in (b)-(c) and (e)-(f) (the black solid envelope that does not exhibit oscillations).

In Fig.~\ref{fig:dynamics} typical examples of the diffraction dynamics are shown in the case of two-photon absorption for self-focusing (left) and self-defocusing (right) nonlinearity. We would like to point out that the intensity patterns for the two opposite signs of the Kerr effect are noticeably different.  We observe that due to the presence of losses, the amount of power transmitted from the left slit is reduced, as compared to the right slit. As a result the envelope of the interference pattern is shifted towards the right. We note that, even at smaller distances (at $z=L_D$) the comparison between theoretical and numerical results is very precise.

\begin{figure}[t]
\centerline{\includegraphics[width=\columnwidth]{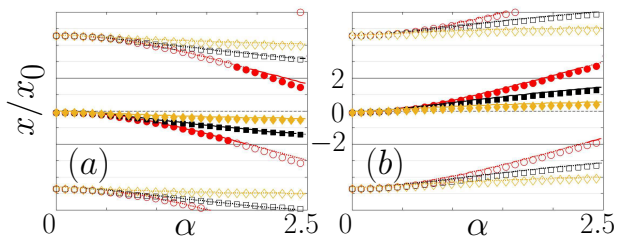}}
\caption{The transverse location of the strongest intensity maxima of the interference pattern as a function of the relative incident amplitude ($\alpha$) for self-focusing (left) and self-defocusing (right) nonlinearity. Filled and hollow symbols indicate the first and secondary intensity peaks, respectively. The reference value of the nonlinear length is $L_{NL,R} =0.025L_D$ and the results for $\sigma^{(2)}=0.2,1,5$, are depicted with (red) circles, (black) squares, and (yellow) diamonds, respectively. The corresponding theoretical results are depicted with curves of the same color. 
\label{fig:centers}}
\end{figure}
We expect that the suggested configuration might be used in nonlinear interferometric measurements. Here we provide a specific example in measuring the nonlinear properties of materials. In particular, it is possible to measure the multi-photon absorption parameters of a medium by directly utilizing Eq.~(\ref{eq:Inl}). We utilize the fact that this equation is independent from the Kerr nonlinear coefficient and depends only on the multiphoton losses inside the medium. A direct fit of the numerical (or theoretical) results to the experimental data at different incident intensities can be utilized to obtain the relevant parameters ($K$ and $L_{MP}^{(K)}$ in our case). Once the multiphoton coefficients are obtained, the next step is to measure the shift in the peaks of the intensity pattern at the observation plane and compare them with the theoretical or even the numerical results. For example, in Fig.~\ref{fig:centers} we depict the location of the strongest intensity peaks of the interference pattern for both self-focusing and self-defocusing nonlinearity, and for three different values of the two-photon absorption coefficient. Since, in the case of self-focusing (self-defocusing) nonlinearity the optical path of the left arm increases (decreases) as the intensity of light increases these peaks translate along the left (right) direction. Note that as the two-photon absorption coefficient increases the losses also increase and, as a result, the phase accumulation and the amount of spatial shifting of the intensity peaks decreases. The comparison of the numerical results with the theoretically predicted maxima is in very good agreement in all of the cases shown in Fig.~\ref{fig:centers}. The comparison is slightly less accurate when the nonlinearity is of the self-focusing type, the two-photon absorption coefficient is weak, and the incident beam intensity is strong. This is attributed to the stronger focusing effects that takes place inside the slab. As we can see, the fringes of the resulting interference pattern are more prominent (the contrast is higher) and regular when the observation plane is shifted to longer distances (compare the cases shown in Fig.-\ref{fig:dynamics} where the observation plane is set to $z_f=3L_D$ and $z_f=L_D$). Thus, selecting a large enough value of the observation plane plays an important role in tracking the intensity maxima of the structure.

For implementation purposes it is interesting to present a specific example in physical units of the nonlinear double-slit configuration presented in this work. Although we use order of magnitude estimations for the material parameters, we have in mind the specific properties of AlGaAs~\cite{ville-apl1993}. Thus, we select a medium with refractive index $n_0=2$, a Kerr nonlinear coefficient $\gamma=10^{-13}$ cm$^2$/W, and a two-photon absorption coefficient $\beta^{(2)}=1$ cm/GW. From these values we obtain the material parameter $\sigma^{(2)}=0.16$ For a square aperture with $x_0=100$ $\mu$m, a laser with wavelength $\lambda=1$ $\mu$m, and a slab length $l_S=L_S/L_D=1/40$ we obtain $L_D=63$ mm and $L_S=1.57$ mm. A maximum phase accumulation $\phi_{NL}=\pi$ is obtained for index contrast $\Delta n_{NL}=\gamma I_0 = 2.06\times10^{-4}$. In the specific example selected, $L_{NL} = 0.385$  mm whereas $L_{MP}^{(2)}=2.42$ mm. For the parameters described above the total power required for both slits is $P = 413$ kW. Note that the parameters presented here are not significantly different from those obtained in~\cite{palto-ol2015} in the absence of losses. This happens mainly because the two-photon absorption length is almost an order of magnitude larger that the nonlinear length.

\section{Conclusions}

We have analyzed a nonlinear interferometric setup based on Young's double-slit experiment where a nonlinear material is placed exactly in front of one of the two slits.
We have examined the effects of Kerr nonlinearity and multi-photon absorption in the resulting interference pattern. The presence of the slab breaks the parity along the $x$-direction as a function of the incident intensity. We have developed a simple theoretical model, based on the modification of the optical path, that is surprisingly accurate in predicting the intensity profile of the main lobes for a wide range of parameters. We discuss about possible applications of our model in measuring the nonlinear properties (Kerr nonlinearity and multiphoton absorption) of different materials. 

An important extension of this work is to consider the case where instead of a laser with continuous wave operation, pulsed laser profiles are used. Such as analysis can rely on averaging the dynamics over the whole pulse as for example in the case of the z-scan method~\cite{sheik-ol1989,sheik-jqe1990}.

\section{Funding Information}
This work was carried out in the framework of the ``Erasmus Mundus NANOPHI project,'' contract number 2013-5659/002-001.

\appendix

\section{Linear losses}
The theoretical model developed in the main text can be generalized even in the case of linear losses. 
The paraxial model still supports exact solutions, however, the resulting expressions become more complicated and less intuitive. Specifically,  
by including linear losses to Eq.~(\ref{eq:paraxial_normalized}) we have 
\begin{multline*}
i\Psi_Z +\frac1{2n_0}\nabla_{X,Y}^2\Psi+
\Gamma\frac{1}{l_{NL}}|\Psi|^2\Psi \\
+i\frac{1}{2l_{L}}\Psi
+i\frac{1}{2l_{MP}^{(K)}}|\Psi|^{2(K-1)}\Psi=0
\end{multline*}
where $L_L$ is the linear loss coefficient. 
Ignoring diffraction and decomposing the wave into constant amplitude and phase we obtain the following expressions for the intensity 
\[
I_{NL} = \frac{e^{-\tfrac{L_S}{L_L}}}{[1+\tfrac{L_L}{L_{MP}^{(K)}}(1-e^{-(K-1)\tfrac{L_S}{L_L}})]^{1/(K-1)}}
\]
and the phase
\[
\phi_{NL} = \Theta(L_S)-\Theta(0)
\]
\begin{multline*}
\Theta(\xi) =  \frac{\Gamma L_{L}}{L_{NL}}e^{-\frac{\Gamma \xi}{L_{L}}} 
{}_2F_1
\left(1,1,\frac{K}{K-1},\frac{e^{-(K-1)\frac{\xi}{L_{L}}}}{1+\tfrac{L_{MP}^{(K)}}{L_{L}}}
\right)
\\
\left(
\frac{L_{MP}^{(K)}+L_{L}-L_{L}e^{-(K-1)\tfrac{\xi}{L_{L}}}}
{(1+\tfrac{L_{L}}{L_{MP}^{(K)}})(L_{MP}^{(K)}+L_{L}(1-e^{-(K-1)\tfrac{\xi}{L_{L}}}))}
\right)^{1/(K-1)}
\end{multline*}
at the output of the slab, where ${}_2F_1$ is the Gauss hypergeometric function. The above formulas for $I_{NL}$ and $\phi_{NL}$ can be substituted to \eqref{eq:fresnel} to obtain the predicted diffraction pattern.

\newcommand{\noopsort[1]}{} \newcommand{\singleletter}[1]{#1}

\end{document}